\documentclass[%
 reprint,
superscriptaddress,
 amsmath,amssymb,
 aps,
pra,
]{revtex4-1}
\usepackage{graphicx}% Include figure files
\usepackage{epstopdf}
\usepackage{dcolumn}% Align table columns on decimal point
\usepackage{bm}% bold math
\usepackage{braket}
 \usepackage{color}

 \definecolor{Revise}{rgb}{0.00,0.00,0.00}
 \definecolor{ReviseA}{rgb}{0.00,0.00,0.00}%
 \definecolor{RA}{rgb}{0.00,0.00,0.00}
 \definecolor{PRA_B}{rgb}{0,0,0}% {0.50,0.00,0.25}

\begin{document}

\preprint{APS/123-QED}

\title{Multiplexing heralded single-photon in orbital angular momentum space}
\author{Shi-long Liu}
 \affiliation{Key Laboratory of Quantum Information, University of Science and Technology of China, Hefei, Anhui 230026, China; and Synergetic Innovation Center of Quantum Information \& Quantum Physics, University of Science and Technology of China, Hefei, Anhui 230026, China
}
\author{Qiang Zhou}
\email{zhouqiang@uestc.edu.cn}
\affiliation{
Institute of Fundamental and Frontier Science, University of Electronic Science and Technology of China, Chengdu 610054, China; and School of Optoelectronic Science and Engineering, University of Electronic Science and Technology of China, Chengdu 610054, China
}%
\affiliation{Key Laboratory of Quantum Information, University of Science and Technology of China, Hefei, Anhui 230026, China; and Synergetic Innovation Center of Quantum Information \& Quantum Physics, University of Science and Technology of China, Hefei, Anhui 230026, China
}%
\author{Zhi-yuan Zhou}%
\email{zyzhouphy@ustc.edu.cn}
\affiliation{Key Laboratory of Quantum Information, University of Science and Technology of China, Hefei, Anhui 230026, China; and Synergetic Innovation Center of Quantum Information \& Quantum Physics, University of Science and Technology of China, Hefei, Anhui 230026, China
}%
\affiliation{
Heilongjiang Provincial Key Laboratory of Quantum Regulation and Control, Wang Da-Heng Collaborative Innovation Center, Harbin University of Science and Technology, Harbin 150080, China.}
\author{Shi-kai Liu}
\affiliation{Key Laboratory of Quantum Information, University of Science and Technology of China, Hefei, Anhui 230026, China; and Synergetic Innovation Center of Quantum Information \& Quantum Physics, University of Science and Technology of China, Hefei, Anhui 230026, China
}%
\author{Yan Li}
\affiliation{Key Laboratory of Quantum Information, University of Science and Technology of China, Hefei, Anhui 230026, China; and Synergetic Innovation Center of Quantum Information \& Quantum Physics, University of Science and Technology of China, Hefei, Anhui 230026, China
}%
\author{Yin-hai Li}
\affiliation{Key Laboratory of Quantum Information, University of Science and Technology of China, Hefei, Anhui 230026, China; and Synergetic Innovation Center of Quantum Information \& Quantum Physics, University of Science and Technology of China, Hefei, Anhui 230026, China
}%
\author{Chen yang}
 \affiliation{Key Laboratory of Quantum Information, University of Science and Technology of China, Hefei, Anhui 230026, China; and Synergetic Innovation Center of Quantum Information \& Quantum Physics, University of Science and Technology of China, Hefei, Anhui 230026, China
}%
\author{Zhao-huai Xu}
\affiliation{Key Laboratory of Quantum Information, University of Science and Technology of China, Hefei, Anhui 230026, China; and Synergetic Innovation Center of Quantum Information \& Quantum Physics, University of Science and Technology of China, Hefei, Anhui 230026, China
}%
\author{Guang-can Guo}
\affiliation{Key Laboratory of Quantum Information, University of Science and Technology of China, Hefei, Anhui 230026, China; and Synergetic Innovation Center of Quantum Information \& Quantum Physics, University of Science and Technology of China, Hefei, Anhui 230026, China
}%
\affiliation{
Institute of Fundamental and Frontier Science, University of Electronic Science and Technology of China, Chengdu 610054, China; and School of Optoelectronic Science and Engineering, University of Electronic Science and Technology of China, Chengdu 610054, China
}%
\affiliation{
Heilongjiang Provincial Key Laboratory of Quantum Regulation and Control, Wang Da-Heng Collaborative Innovation Center, Harbin University of Science and Technology, Harbin 150080, China.}
\author{Baosen Shi}
\email{drshi@ustc.edu.cn}
\affiliation{Key Laboratory of Quantum Information, University of Science and Technology of China, Hefei, Anhui 230026, China; and Synergetic Innovation Center of Quantum Information \& Quantum Physics, University of Science and Technology of China, Hefei, Anhui 230026, China
}%
\affiliation{
Heilongjiang Provincial Key Laboratory of Quantum Regulation and Control, Wang Da-Heng Collaborative Innovation Center, Harbin University of Science and Technology, Harbin 150080, China.}
\date{\today}% It is always \today, today,
\begin{abstract}
Heralded single-photon source (HSPS) with competitive single photon purity and indistinguishability has become an essential resource for photonic quantum information processing. Here, for the first time, we proposed a theoretical regime to enhance heralded single-photons generation by multiplexing the degree of the freedom of orbital angular momentum (OAM) of down-converted entangled photon pairs emitted from a nonlinear crystal. Experimentally, a proof-of-principle experiment has been performed  through multiplexing three OAM modes. We achieve a 47$\%$ enhancement in single photon rate. A second-order autocorrelation function $g^{(2)}(0)<0.5$ ensures our multiplexed heralded single photons with good single photon purity. We further indicate that an OAM-multiplexed HSPS with high quality can be constructed by generating higher dimensional entangled state and sorting them with high efficiency in OAM space. Our avenue may approach a good HSPS with the deterministic property.
\end{abstract}
\pacs{Valid PACS appear here}% PACS, the Physics and Astronomy
\maketitle
\section{Introduction}
Perfect single photon source, i.e., emitting indistinguishable single photon on-demand, is a fundamental element for realizing the quantum information processing, such as linear quantum computations \cite{gisin2002quantum,knill2001scheme,ladd2010quantum}, cryptography \cite{bennett2014quantum}, and metrology \cite{giovannetti2006quantum}. In {\color{Revise}{the last several decades}}, many platforms have been made to develop such a source \cite{aharonovich2016solid}, including quantum dots \cite{ding2016demand,somaschi2016near}, color centers in diamond or 2D materials \cite{kurtsiefer2000stable,aharonovich2016solid}, and molecules or atoms \cite{lounis2000single,chou2004single}. Furthermore, we intend to obtain high-quality heralded single-photon source (HSPS) by multiplexing certain degrees of the freedom (DOF) of photon \cite{migdall2002tailoring,pittman2002single}.

For a multiplexed HSPS, an essential resource is entangled photon pairs emitted from nonlinear materials, for instance, processes of spontaneous parametric down-conversion (SPDC) and spontaneous four-wave mixing. Many multiplexed HSPSs with high indistinguishability have been demonstrated in various DOF  of photon \cite{francis2017active}, such as in spatial \cite{collins2013integrated,francis2016all,mazzarella2013asymmetric}, temporal \cite{kaneda2015time,xiong2016active}, polarization \cite{ma2011experimental}, spectral \cite{puigibert2017heralded,joshi2018frequency}, and in both spatial and temporal
\cite{mendoza2016active,kaneda2018high}. For the spatial multiplexing HSPS, the strategy is to multiplex photon pairs in different spatial paths into a single path with the help of the optical switch, where one used to employ multiple basic units to generate photon pairs, i.e., arrayed silicon waveguides \cite{collins2013integrated}, bulk crystals \cite{migdall2002tailoring} and fibres \cite{francis2016all}. The higher rate of the heralded single photon, the more basic units one needs, which results in poor scalability. Temporal multiplexing technology with one basic unit has good scalability. However, a successful manipulation requires the complex timing sequence \cite{kaneda2015time} and sacrifices the repetition rate. Another scheme with one basic unit is worked in spectral mode \cite{puigibert2017heralded,joshi2018frequency}. It is a very promising scheme to develop multiplexed HSPS approaching to a deterministic single photon source, while manipulating spectral modes needs robust optical filtering otherwise very noisy \cite{joshi2018frequency}, or the performance of frequency shifting devices limiting the number of modes \cite{puigibert2017heralded}. Therefore, it is a valuable direction to explore a unique high-performance HSPS with a single basic unit multiplexed in new DOF of photon.

Recently, orbital angular momentum (OAM), a remarkable DOF of photon with infinite dimensions, has obtained increasing attentions and rapid developments \cite{Allen1992,Yao2011a,erhard2018twisted,wang2012terabit}. High-dimensional photonic state encoded with vortex wavefront ($e^{iL\phi}$) can increase the information capacity, which is natural and quite essential for scalability in quantum information applications. {\color{Revise}{Much progress of high-dimensional quantum state has been achieved \cite{Yao2011a,erhard2018twisted}, such as realizing or simulate multilevel quantum systems \cite{Mair2001,bouchard2018measuring,liu2018generation,liu2019violation,doi:10.1080/00107514.2019.1580433}, generating a high-dimensional maximally entangled state \cite{Dada2011,franke2004uncertainty,malik2016multi,liu2018coherent,kovlakov2018quantum}, and exploring the high-dimensional quantum computations, i.e., high-dimensional single-photon gate \cite{Babazadeh2017} and Bell basis \cite{Wang2017}.
Towards generating a high-dimensional entangled state in OAM space for multiplexed HSPS, the most common method is to engineer down concerted photon pairs emitted from a nonlinear crystal through the process of SPDC \cite{Mair2001}. According to the OAM conservation between signal, idler and pump photons, i.e. $L_p=L_s+L_i$, the output state can be written as $\sum_{L=-\infty}^{\infty}c_L\ket{L}_s\ket{-L}_i$, where the distribution of amplitudes ${c_L}$ associates with the spiral bandwidth of entangled two-photon state, which usually rapidly decreases as the increase of topological number of OAM \cite{Torres2003}. Previous works on multiplexed HSPS always focus on parts of zero state, or Gaussian beam, and neglect the occupations of high orders in OAM space. Certainly, we cannot ignore the amplitude of $c_L (|L|\neq0)$, for example, the amplitudes of $c_{\pm1}$ near the zero order have a considerable value \cite{Torres2003}. Therefore a high-efficiency and well scalable multiplexed HSPS would be developed if the high-order OAM parts employed in such system.}}

{\color{ReviseA}{In this letter, we propose and experimentally investigate a multiplexed HSPS in the OAM DOF of the photon. First, we describe the fundamental principle to prove that the single photon generations can be enhanced by multiplexing more OAM entangled photon pairs in the SPDC process. Second, an ideal experimental regime is proposed by using a feed-forward control system. Then, a proof-of-principle HSPS is demonstrated by multiplexing  three OAM modes: $c_{-1}\ket{-1}_s\ket{1}_i+c_0\ket{0}_s\ket{0}_i+c_1\ket{1}_s\ket{-1}_i$.
Finally, we make a discussion about the OAM-HSPS. In addition, some OAM mode sorting methods are posted in the appendix.
In our experiment, the enhancement of 47\% is achieved compared to the only one Gaussian mode. To verify that the OAM-HSPS is indeed in the single-photon regime, we carry out the Hanbury-Brown and Twiss (HBT) experiment to measure the second-order autocorrelation function $g^{(2)}(\tau)$. A smaller $g^{(2)}(0)$ indicates a higher single photon purity \cite{brown1956correlation,bocquillon2009coherence}. We obtained the value of $g^{(2)}(0)=0.097\pm0.011$ for OAM-HSPS. The results clearly illustrate that the OAM-HSPS is of good quality and is a promising avenue for single photon-source. The only notable disadvantage is the loss of the OAM-sorter, but it is an improvable factor during the multiplexing. In fact, it has been a valuable direction to reduce the losses and sorting-visibility in sorting OAM modes \cite{Yao2011a}. We make a detailed discussion in the appendix.}}
%\section{Principle}
{\color{PRA_B}{
\section{Enhancement of single-photon generations via multiplexing OAM modes}
 In the process of spontaneous parametric down conversion (SPDC), the quantum fluctuations would be neglected if the pump field is undepleted. And thus one can describe the SPDC as a process of the optical parametric amplification (OPA) \cite{christ2011probing}, where the down-conversion state is given in the interaction picture:
 \begin{equation}
\begin{array}{l}
{\left| \psi  \right\rangle _{SPDC}} = exp\left[ { - \frac{i}{\hbar }\left( {B\sum\limits_{k,l} {\int {\int d } } {\omega _s}d{\omega _i}{f_{k,l}}({\omega _s},{\omega _i}) \times } \right.} \right.\\
\;\;\;\;\;\;\;\;\;\;\;\;\;\;\;\;\;\left. {\left. {a_k^{\left( s \right)\dag }({\omega _s})a_l^{\left( i \right)\dag }({\omega _i}) + H.c.} \right)} \right]\;\left| 0 \right\rangle
\end{array}
 \end{equation}
Where $a_k^{(s)\dag}({\omega _s})\;,a_l^{(i)\dag}({\omega _i})$ represent the photon creating operators in spatial $k$ and $l$ with frequency ${\omega _s}$ and ${\omega _i}$; $\ket{0}$ is the vacuum state;  ${f_{k,l}}({\omega _s},{\omega _i})$ describes the correction function in both spatial and spectrum degree of freedom (DOF) of down converted photon, which is depended on both nonlinear optical material and pump structure. If one considers the only one DOF of the photon pairs, i.e., the spatial mode, the state can be written as \cite{christ2012limits}:
\begin{equation}
  \left| \psi  \right\rangle {\rm{ = }}\mathop  \otimes \limits_k \sqrt {1 - \gamma _k^2} \sum\limits_{n = 0}^{n = \infty } {\gamma _k^2\left| {n_k^{(s)},n_k^{(i)}} \right\rangle } \
\end{equation}
 where ${\gamma _k}( = P{\tau _k})$ describes the interaction parameter including pump power $P$ and the interaction constant ${\tau _k}$ of the nonlinear material \cite{broome2011reducing,jin2015efficient}; $\left| {n_k^{(s)},n_k^{(i)}} \right\rangle $ represents $n$ pair photon state containing $n$ photons in $k$ mode.  The probability of a successful heralding event containing two-photon pairs in $k$ mode is calculated by:
 \begin{equation}
 {{\Pr}_{k}}(1,1) = \left\langle \psi  \right|\left| {1_k^{(s)},1_k^{(i)}} \right\rangle \left\langle {1_k^{(s)},1_k^{(i)}} \right|\left| \psi  \right\rangle
 \end{equation}
 Base on the Eq. (3), one can calculate the single counts $S{C_k}(1,1) = {R_0} \times {\eta _T} \times {\Pr _k}(1,1)$ and coincidence counts
 $C{C_k}(1,1) = {R_0} \times \eta _T^2 \times {\Pr _k}(1,1)$, respectively \cite{jin2015efficient}. When $k=0$, the prepared output state corresponds to the familiar forms of ${\left| \psi  \right\rangle _{k = 0}}{\rm{ = }}\sqrt {1 - {\gamma ^2}} \sum\limits_{n = 0}^\infty  {{\gamma ^2}\left| {{n^{(s)}},{n^{(i)}}} \right\rangle } $.
 One can multiplex certain DOFs of photon to achieve an HSPS close to ideal single photon source, i.e., temporal modes \cite{kaneda2015time}, polarization \cite{ma2011experimental}, and spatial paths \cite{collins2013integrated,mendoza2016active}. However, most of case,
there are some occupations in high-order spatial mode \cite{bennink2010optimal,miatto2011full}. In that case, the probability of successful heralding event ${\Pr _k}(1,1)$ and coincidence $C{C_k}(1,1)$ will increase linearly if we can successfully multiplex many $k$ modes simultaneously. Very recently, the frequency-multiplexing heralded sources have been demonstrated in some works, i.e., Refs \cite{puigibert2017heralded,joshi2018frequency}. By considering the only situation of two photon pairs, the state in Eq. 2 can be written in:
 \begin{equation}
   \left| \psi  \right\rangle  = \mathop  \otimes \limits_k \sqrt {1 - \gamma _k^2} \sum_{k } {\gamma _k^2\left| {1_k^{(s)},1_k^{(i)}} \right\rangle }
 \end{equation}
 The state can be Schmidt decomposition in Laguerre Gaussian modes \cite{miatto2011full}:
 \begin{equation}
  \begin{array}{l}
\left| \psi  \right\rangle {\rm{ = }}\sum\limits_{{l_s}{p_s}} {\sum\limits_{{l_i}{p_i}} {c_{{p_s},{p_i}}^{{l_s},{l_i}}} } \left| {{1_{{l_s},{p_s}}}} \right\rangle \left| {{1_{{l_i},{p_i}}}} \right\rangle \\
\;\;\;\;\;{\rm{ = }}\sum\limits_{{l_s}{p_s}} {\sum\limits_{{l_i}{p_i}} {c_{{p_s},{p_i}}^{{l_s},{l_i}}} } \left| {{l_s},{p_s}} \right\rangle \left| {{l_i},{p_i}} \right\rangle
\end{array}
    \end{equation}
 Where the coefficient $c_{{p_s},{p_i}}^{{l_s},{l_i}}{\rm{ = }}\left\langle {{l_s},{p_s};{l_i},{p_i}} \right.\left| {{\psi}} \right\rangle$ gives the amplitude probability finding the photon pairs in corresponding spatial mode \cite{miatto2011full,liu2018coherent}. If we don't consider the radial mode ($p=0$) and successfully multiplex all of the OAM modes, the probability of a successful heralding event can be calculated:
 \begin{equation}
   \Pr ({l_s},{l_i}) = \sum\limits_{l,k =  - \infty } {{c_l}{c_k} = } \sum\limits_{l - \infty } {c_l^2}
 \end{equation}
Where $c_0$ is the amplitude probability for Gaussian mode, no multiplexing situation. The single counts and coincidence counts can be calculated in following \cite{jin2015efficient}:
\[S{C_s}(l) = {R_0} \times {\eta _s} \times \Pr ({l_s}) = {R_0} \times {\eta _T} \times \sum\limits_l {{\eta _l}c_l^2} \]
\[S{C_i}(l) = {R_0} \times {\eta _i} \times \Pr ({l_i}) = {R_0} \times {\eta _T} \times \sum\limits_l {{\eta _l}c_l^2}  \]
\[CC(k) = {R_0} \times \eta _T^2 \times \Pr ({l_s}) = {R_0} \times \eta _T^2 \times \sum\limits_l {\eta _l^2c_l^2} \]
where $1 - {\eta _T}$ represents the independent losses for all the OAM modes; $1 - {\eta _l}$ shows the losses for different OAM modes, i.e., the mode sorting efficiencies; $R_0$ is a constant. One can define the enhancement for OAM- HSPS in the probability of a successful heralding even (coincident counts) with respect to the only Gaussian mode:
\begin{equation}
  \Gamma  = \frac{{CC\left( k \right)}}{{CC(k = 0)}} = \frac{{{R_0} \times \eta _T^2 \times \sum\limits_l {\eta _l^2c_l^2} }}{{{R_0} \times \eta _T^2 \times c_{l = 0}^2}} = \frac{{\sum\limits_l {\eta _l^2c_l^2} }}{{c_{l = 0}^2}}
\end{equation}

\begin{figure}
  \centering
  \includegraphics[width=8cm]{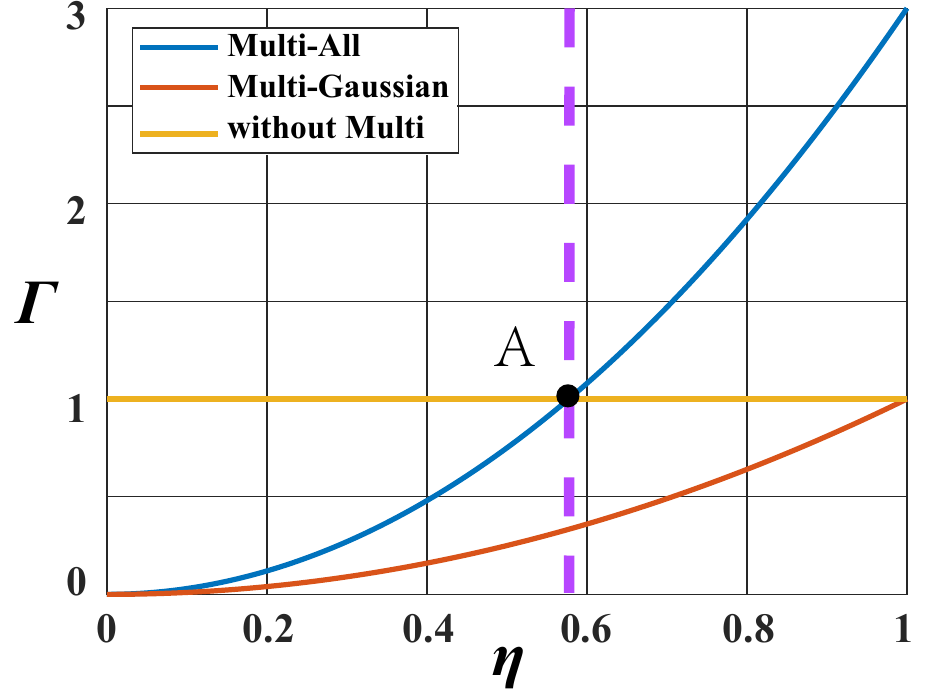}
  \caption{The enhancement for a three-dimensional maximally entangled state. The x-axis represents the losses dependent parameter; the y-axis is the enhancement defined in Eq. 7. Here, Multi-All (blue and upper line) and Multi-Gaussian (red and lower line) represent the enhancements with three OAM modes and with one of modes, i.e., Gaussian mode, under the multiplexing setup;  The horizontal yellow line is the Gaussian mode without the multiplexing setup.  The vertical dashed line represents the enhancement of the multiplexing source is equal to only one Gaussian mode (marked on point A).}\label{4}
\end{figure}

  In the Eq. 7, one can find that the more OAM modes are multiplexed; the higher enhancement $\Gamma$ is obtained. The enhancement is strongly dependent on losses $1-\eta_l$ in mode sorting. In Fig. 1, we plot the relationship between losses and enhancements, where we assume that the input is a three-dimensional maximum entangled state: $\ket{-1}+\ket{0}+\ket{1}$, and the losses for all the OAM mode are equally. In Fig. 1, the blue (upper) line is the enhancement in photon generation  for
 multiplexing source, and the red (lower) line is Gaussian mode including losses from the mode sorting and collection etc. One can find that the multiplexing source will get the enhancements compared without multiplexing.  The yellow (horizontal) line is the situation of without multiplexing, only one Gaussian mode, where we don't consider the losses. One should note that there is a limited value in losses marked by purple (vertical) dashed line, which illustrates the multiplexing regime is better than without multiplexing.  For a d-dimensional maximum state, it is $1/\sqrt{d}$.

\section{An ideal regime to realize OAM mode multiplexing}
\begin{figure}
  \centering
  \includegraphics[width=8cm]{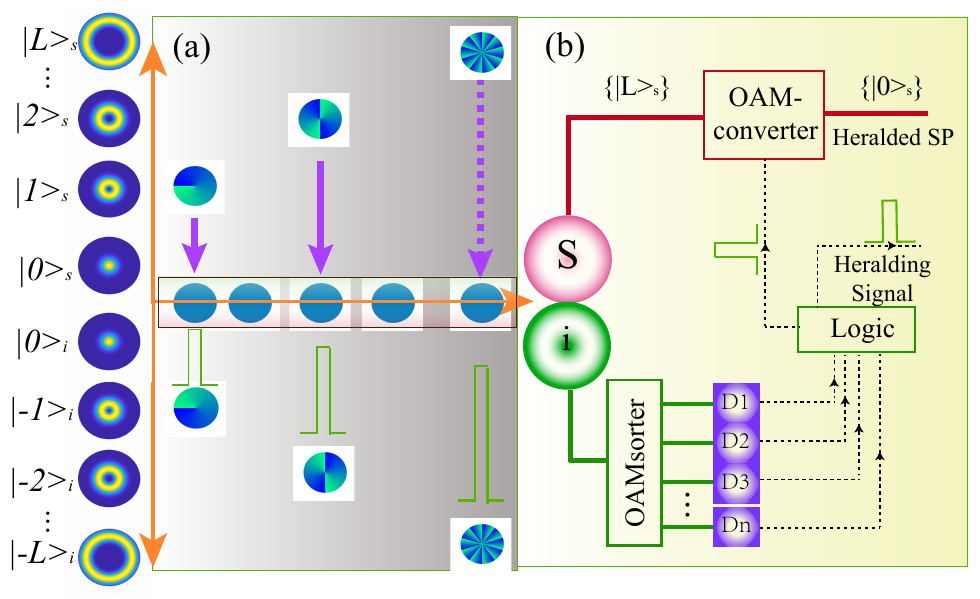}
  \caption{The principle of the OAM-HSPS. a, the schematic of multiplexing OAM modes. b, the simple regime for OAM-HSPS. High-dimensional entangled photon pairs (signal and idler photons) are from the process of SPDC. The state $\ket{L}_s$ of signal photon will be converted instantly to a signal photon without vortex phase triggered by the heralding signal. }\label{0}
\end{figure}
{\color{PRA_B}{In Fig. 2, we propose an ideal regime to realize an OAM-HSPS. The element source is the entangled photon pairs from the process of SPDC. By considering the only azimuthal variable of $L$, the state can be written as $\sum_{L=-\infty}^{\infty}c_L\ket{L}_s\ket{-L}_i$\cite{Torres2003}. Here,the state $\ket{L}_s\ket{-L}_i$ represents one signal photon in the spatial mode of $L_s$ and another idler photon in the spatial mode of $-L_i$. ${c_L}$ represents the amplitudes for each OAM entangled photon pairs, sometimes associated with Schmidt number \cite{Torres2003,miatto2011full}. In general, the Schmidt number is not equal to 1, and the single mode condition is difficult to meet in experiment \cite{Torres2003}. Therefore, multiplexing high-order OAM modes is beneficial to increase the probability of a successful heralding event.
 \begin{figure}
   \centering
   \includegraphics[width=8cm]{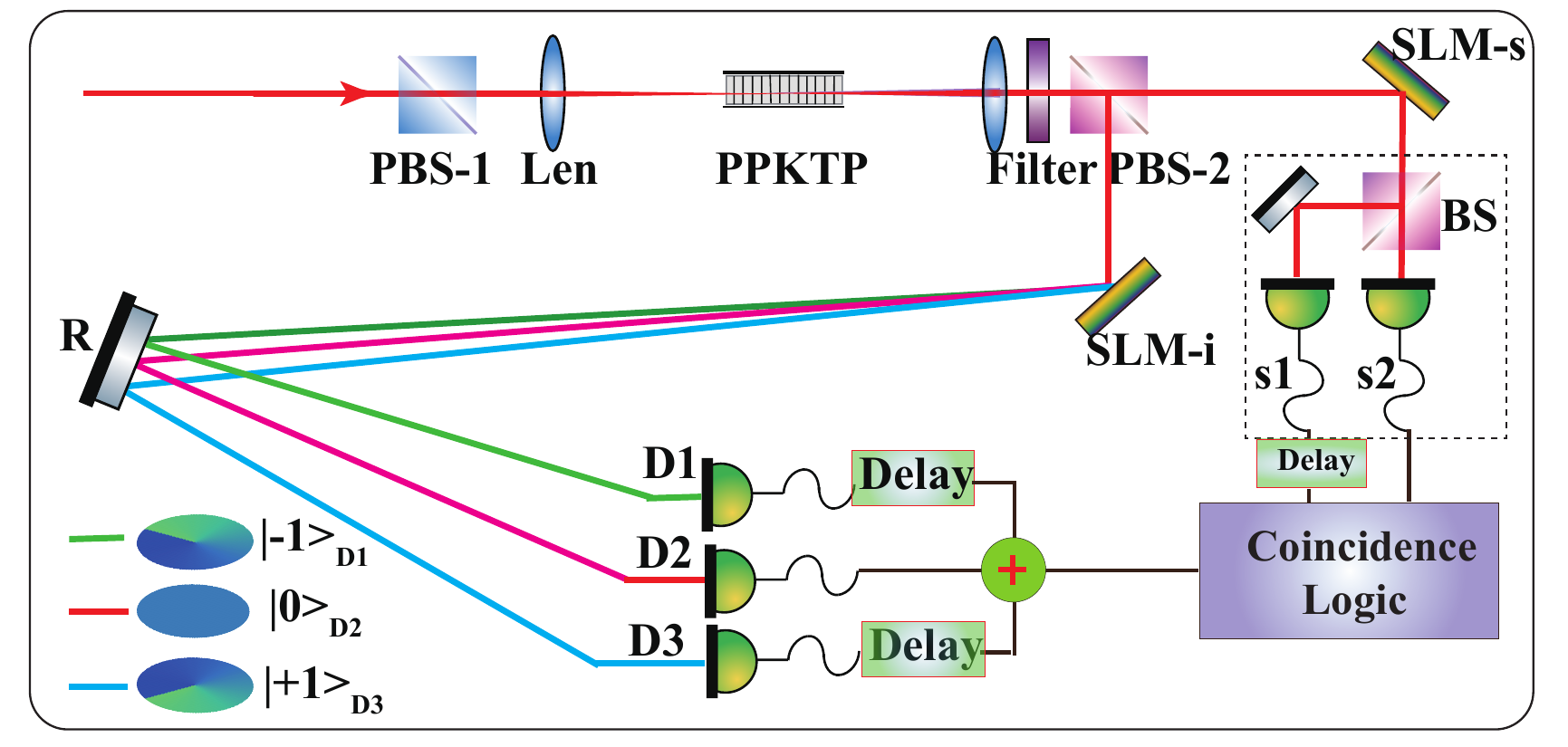}
   \caption{Schematics for OAM-HSPS. A narrow bandwidth 780 nm laser beam (Ti: sapphire, MBR110, Coherent) pumps a type-II Periodically Poled Potassium Titanyl Phosphate (PPKTP,$1\times2\times10$  $mm^3$). The beam of signal and idler photons in the center of PPKTP are exactly imaged on the planes of two SLMs, i.e. SLM-s and SLM-i, by using a convex lens. Signal and idler photons are separated by a polarization beam splitter (PBS). Both SLM-s and SLM-i are encoded with opposite phases with blazed grating.}\label{1}
 \end{figure}
Fig. 2(b) shows the principle of the feed-forward controlling regime, where the idler photons are heralding photons, and the signal photons are heralded ones. The idler photons are sorted to one of the paths based on the value of $-L_i$ through an OAM-sorter. Later, the idler photon $\ket{-L}_i$ is detected by a single photon detector $(D_n)$ for generating a heralding electronic signal. In the meantime, the entangled signal photon $\ket{L}_s$ arrive at an OAM-converter; after passing through the OAM-converter, the vortex phase $exp(iL_s\phi)$ of the signal photon is converted instantly to a plane phase $(L_s=0)$ for uses. Fig. 2(a) shows the simplified diagram of the OAM-HSPS. By adding more and more OAM modes, the heralded photons will yield a considerable improvement compared with  only one Gaussian mode.

For the scheme in Fig. 2 with an actively feed-forward element, i.e., a high-speed SLM, there is only reflection loss in mode sorting, and thus we can realize the maximum enhancement in Eq. 7.  In the overall process of the multiplexing, two essential elements are needed, i.e., an OAM-sorter with high-efficiency and an OAM-converter with high-speed. Many OAM-sorters have been demonstrated, such as Mach-Zehnder interferometers-based \cite{leach2002measuring} and spatial light modulators-based \cite{Yao2011a,berkhout2010efficient}. In interferometric method of Ref. \cite{leach2002measuring}, we have to use numerous optical elements,i.e., dove lens, for separating three OAM modes, which is not suitable in our demonstration. By performing a Cartesian to log-polar coordinate transformation \cite{berkhout2010efficient}, we can efficiently sort several OAM modes. However, we need even three SLMs and the global sorting efficiency will be declined. A high-quality OAM-sorter with sorting as many as possible modes is still missing. In the appendix, we list some effective methods to sort OAM photon. Meanwhile, the performance of OAM-converter is also far from perfect. For instance, the input frame rate of commercial SLMs is limited to several hundred Hertz.

The feed-forward controlling scheme needs an ideal OAM sorter, while none of which are of high-speed and high-efficiency. In our proof-of-principle experiment, we employ two SLMs to sort OAM mode using Forked diffraction blazing to indirectly prove that the enhancement in Eq. 7. Here, all of the mode is unconditionally converted into Gaussian mode with the conversion efficiency $1/N$.  Although the employed OAM-sorters and -converters are not ideal, the demonstration can be as a good proof-of-principle experiment for OAM-HSPS in our work. }}
\begin{figure}
   \centering
   \includegraphics[width=8.5cm]{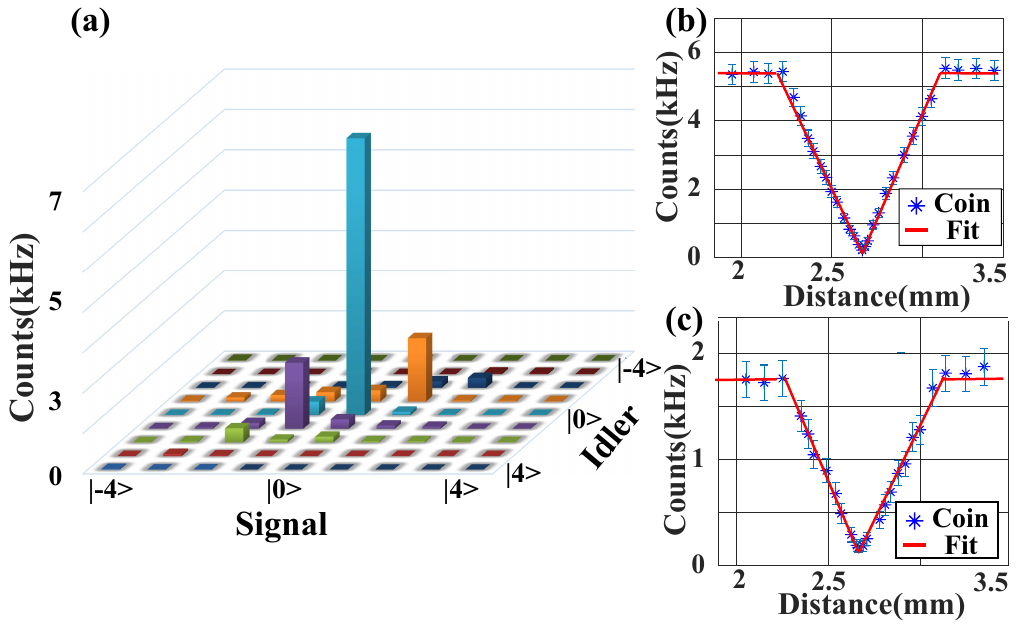}%{Fig_SSHOM.pdf}
   \caption{The spatial distributions and HOM interference curves of down-converted photon pairs. a: The  coincidence are measured for OAM eigenstates from $\ket{-4}$ to $\ket{4}$. The pump is a Gaussian beam at 780 nm, two SLMs load the corresponding OAM eigenstates of the pure phase. b and c: The HOM interference curves for down-conversion states of $\ket{0}_s\ket{0}_i$ and $\ket{1}_s\ket{-1}_i$. Each data of coincidence is recorded in one second.
   }\label{2}
 \end{figure}
 }}
\section{The setup for a proof-of-principle HSPS}
The experimental setup is presented in Fig. 3. Photon pairs occupied in three OAM modes are utilized, i.e., $c_{-1}\ket{-1}_s\ket{1}_i, c_0\ket{0}_s\ket{0}_i, c_1\ket{1}_s\ket{-1}_i$. For the heralding side, the OAM-sorter, including a SLM-i, is used to convert three OAM modes into three different spatial Gaussian modes. After the mode sorting, the modes $\ket{-1}_i$, $\ket{0}_i$ and $\ket{1}_i$ are detected by three single photon detectors, i.e., D1, D2, and D3, respectively. The OAM mode-sorting is realized by loading a complex phase of  $Arg(e^{1\times\psi})e^{ikd_{-1}}+Arg(e^{0\times\psi})e^{ikd_{0}}+Arg(e^{-1\times\psi})e^{ikd_{1}}$ onto the SLM-i, where the phase of $Arg(e^{-L_i\times\psi})$ is used to flat the vortex phase of the input photon; $e^{ikd_{L_i}}$ represents a displacement operation for the OAM mode of $\ket{L_i}$ \cite{gibson2004free,li2015simultaneous}. What we need to pay attention to is the above method exist an efficiency of $1/3$ for one of input modes, for example, the state will be $\ket{0}_{D1}+\ket{-1}_{D2}+\ket{-2}_{D3}$ if the input state is $\ket{-1}_{i}$. For the heralded side, we employ a SLM-s to realize the function of the OAM-converter, where the acquired phase is similar to the situation of OAM-sorter beside the displacement operation, $d_{L_s}=0$. After passing through the OAM-converter, all of the modes ${\ket{L}_s}$ of signal photon is converted to ${\ket{0}_s}$ with a probability of 1/3, and mode ${\ket{0}_s}$ is filtered and collected by using a single mode fiber.

{\color{RA}{In our proposal, we sort the heralding photon to different ports based on the corresponding OAM, e.g. D1, D2  in Fig. 2 (OAM sorter). Therefore, this type of proposal regime avoids the continually switching SLMs. Actually, we can adjust the time sequence of the idler photon $\ket{L}_i$ to demonstrate equally the enhancement in the original protocol (see Fig. 3).}}
 \begin{figure*}[tbp]
        \centering
        \includegraphics[width=16cm]{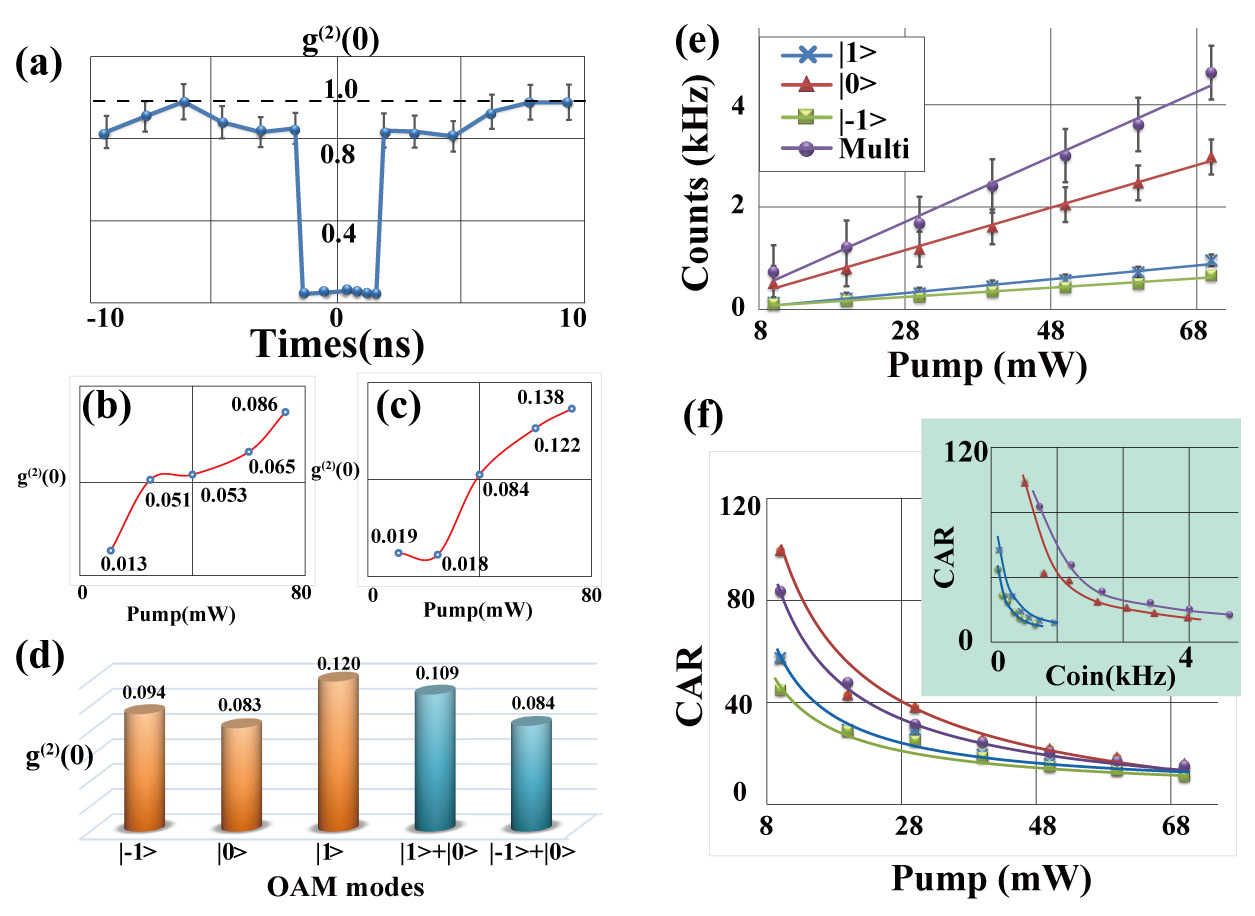}
        \caption{The results of second-order auto correlation function $g^{(2)}(\tau)$ and rates of the single and multiplexing OAM-HSPS. a: The distributions $g^{(2)}(\tau)$ of $\ket{0}_s/\ket{0}_i$;{\color{Revise}{where error bars are estimated using Poisson statistics; the measured average $g^{(2)}(0)=0.048\pm0.003$}}. b and c: The relationship between the value of  $g^{(2)}(0)$ and pump power for two OAM modes of $\ket{-1}+\ket{0}$ and $\ket{0}+\ket{1}$, respectively.
         d: The average $g^{(2)}(0)$ for single or OAM-HSPS. The coincidences during the measurement of $g^{(2)}(\tau)$  are recorded over 40 seconds, and the pump is the Gaussian beam with the power of 77 mW. {\color{Revise}{ The average measured $g^{(2)}(0)=0.094\pm0.017 ,0.083\pm0.008 ,0.120\pm0.019,0.109\pm0.012$,and $0.084\pm0.009$ for single and multi modes.}}
        e: The heralded single-photon rates from OAM modes of $\ket{-1}$(green square dot), $\ket{0}$(red triangle dot), $\ket{1}$(blue cross dot), and after multiplexing (purple circle  dot) are plotted as a range of input powers.
       {\color{ReviseA}{f: The CAR obtained from the  same conditions of a, where a subfigure is inserted into the top right Fig. 5(f) to show the CAR versus coincidence counts.}} Lines are linear fits to the data. In Fig. 5(e) and (f), the coincidence counts are presented in unit time (1 second), while the coincidences in the measurement are recorded 20 seconds. {\color{PRA_B}{In Fig. 5(e) and (f), the coincidence in Gaussian mode $\ket{0}$ are recorded under the situation of multiplexing, which includes the losses in mode sorting. The comparison is like the upper blue and lower red lines in Fig. 1.}} }\label{3}
      \end{figure*}
 \section{Results of enhancements in photon generations}
     Considering three OAM entangled modes $c_{-1}\ket{-1}_s\ket{1}_i+c_0\ket{0}_s\ket{0}_i+c_1\ket{1}_s\ket{-1}_i$,
     the coincidence rate for each OAM entangled mode is proportional to the coefficient $c_0^2$ and $c_{\pm1}^2$. By multiplexing three OAM modes, the enhancement can be improved of $1+2\xi^2$ in contrast to a single Gaussian mode (see supplementary material), where $\xi=|c_1|/|c_0|$ represents the scale of enhancement, for example, for the case with only Gaussian mode\cite{bennink2010optimal,jin2019theoretical}, there is no enhancement due to $\xi=0$;  for the three-dimensional maximum entanglement state \cite{liu2018coherent}, all the amplitudes is equally $(|c_{\pm1}|=|c_{0}|)$, and the enhancement is up to 3; higher dimension corresponds to a higher rate of HSPS. In our experiment, the value of $2\xi^2$ is 0.446, which can be measured by the spatial distributions of down-converted photons. Fig. 4(a) shows the results for the spatial distributions of down-converted photons in OAM space, where x(y)-axis represents the pure OAM eigen states from -4 to 4; z axis is the coincidence recorded in one second.

{\color{ReviseA}{Firstly, we check the  indistinguishability  between different OAM modes in frequency domain. The components in the dashed box, using to test the second-order autocorrelation function as shown in Fig. 3, are removed. A  fiber beam splitter is added into two paths of both signal and idler, where the tunable positioning systems are inserted to adjust the optical path difference \cite{zhou2017superresolving}. By using a SLM,  the vortex photon with non-Gaussian transverse mode structure is converted into the Gaussian mode. The results for Gaussian and OAM modes are depicted in Fig. 4b and Fig. 4c. The error bars were standard deviations, assuming all the coincidences follows Poisson's distributions. Based on the fitting curves, we obtained the visibilities  $95.38\%\pm0.7\%$ and $86.84\%\pm1.1\%$ for the Gaussian ($\ket{0}_s \ket{0}_s$) and OAM modes($\ket{1}_s\ket{-1}_i$), respectively. Because of the crosstalk and imperfect overlap between the phase in SLM and actual distributions of photons, the visibility of $\ket{1}_s\ket{-1}_i$ is a bit low than the Gaussian mode.  The inferred spectral bandwidths \cite{ou2007multi,li2015cw} for $\ket{0}_s\ket{0}_i$  and $\ket{1}_s\ket{-1}_i$  are the same value of 2.67nm, which illustrate that they are identical in the spectral domain. If we flat the vortex phase of nonzero OAM modes simultaneously, they will be identical in temporal, spatial and spectral domains, which ensures that we can construct an HSPS by multiplexing OAM modes.}}

Then, to verify that the output OAM-HSPS maintains the single-photon characters, we measured the second-order autocorrelation function $g^{(2)}(\tau)=C_{s_1,s_2,i}(\tau)\cdot C_{i}/C_{i,s_1}(\tau)\cdot C_{i,s_2}(0)$ of single and after multiplexing sources.  Here, $C_{s_1,s_2,i}(\tau)$ is the three-fold coincidence between the signal of $s_1$ and $s_2 $ separated by a beam splitter (the dashed box in Fig. 3) and the idler of $i$; $C_{i,s_1}(\tau)$ and $C_{i,s_2}(0)$ represent two-fold coincidences between idler and $s_1$,  and $s_2$, respectively; $C_{i}$ is the number of clicks of idler, the heralding counts. $\tau$ is the delay between heralded $s_1$ and heralding $i$, which is usually on a magnitude of nanoseconds \cite{bocquillon2009coherence}. To measure $g^{(2)}(\tau)$, we employ the most common method, HBT setups \cite{brown1956correlation,jin2015efficient}. In our setups, the components in the dashed box shown in Fig. 3 are the HBT setups for measuring the property of the heralded photons. First, we measure the $g^{(2)}(\tau)$ for the single OAM modes, where the phases encoded in SLM-s/SLM-i are $\ket{1}_s/\ket{-1}_i$, $\ket{0}_s/\ket{0}_i$, and  $\ket{-1}_s/\ket{1}_i$, respectively. The obtained $g^{(2)}(0)$ is shown by orange coloured bars in the Fig. 5(c), and the overall measurements of $g^{(2)}(\tau)$ for $\ket{0}_s/\ket{0}_i$ is plotted in the Fig. 5(a). Here, the x-axis is the delays and the coincidence windows is 1.6 ns. Then, we measure the $g^{(2)}(0)$ of the OAM-HSPS.  Here, the phases for two SLMs are $\ket{1}_s+\ket{0}_s+\ket{-1}_s$ and $\ket{-1}_ie^{ikd_{-1}}+\ket{0}_ie^{ikd_{0}}$ or $\ket{0}_ie^{ikd_{0}}+\ket{1}_ie^{ikd_{1}}$. The corresponding values of $g^{(2)}(0)$ are plotted by blue coloured bars in Fig. 5(d). Because of the different collected efficiency between mode of $\ket{1}$ and $\ket{-1}$ during the measurement, the measured $g^{(2)}(0)$ of  $\ket{1}+\ket{0}$ is a bit higher than the situation of  $\ket{-1}+\ket{0}$. Nevertheless, all the measured values of $g^{(2)}(0)$ are around 0.1, which shows the OAM-HSPS with good single photon purity and close to true single photon source. Also, we measure the value of $g^{(2)}(0)$ of OAM-HSPS with the increase of pump power, which are inserted below Fig. 5(a), where the OAM-HSPS with two OAM modes, $\ket{-1}+\ket{0}$, is the left one (Fig. 5(b)), and with $\ket{1}+\ket{0}$ is the right one (Fig. 5(c)), respectively.

Finally, we measure the OAM-HSPS rates with various pump powers. These results are plotted in Fig. 5(e), where green square), red triangle, and blue cross dots show the coincidence with the modes of $\ket{-1}_i\ket{1}_s$, $\ket{0}_i\ket{0}_s$, and $\ket{1}_i\ket{-1}_s$, respectively; the purple circle one gives the result of OAM-HSPS after multiplexing.  Based on four fitting curves, we can infer the enhancement of HSPS with three OAM modes of 1.47, or 47\% increase, which is lower than the theories where the enhancement should be 1.89 for our system($\xi^2=0.446$). The main reason is that the asymmetrical losses between Gaussian and high-order modes during the projection measurements. {\color{PRA_B}{One should note that this enhancement is compared with multiplexing three OAM modes and only one Gaussian mode, like the lines of blue and red lines in Fig.1. For such a state, the value of limited value (like point A in Fig. 1) is 0.83 ($\sqrt{1/(1+2\eta)}$) beyond the only Gaussian mode without multiplexing}}.

 Furthermore, we measured the coincidence-to-accidental coincidence ratio (CAR). In the ideal case, the CAR ($\approx \Pr (n = 1)/\Pr (n = 2)$) represents the ratio of probability between two- and four- photon pairs, which is decreased as the pump power\cite{puigibert2017heralded,collins2013integrated}. The corresponding CAR is depicted in Fig. 5(f), where the colors and shapes of date represent the same meaning as in Fig. 5(e). Similar to the result of another multiplexing systems \cite{collins2013integrated,francis2016all}, the CAR has a rapid decrease as the increase of pump power in our systems. {\color{ReviseA}{For studying the CAR versus coincidence counts, we also insert a subfigure into the top right Fig. 5(f). The results illustrate a fact that the CAR of multiplexed HSPS does improve under the same coincidence.

{\color{RA}We did not directly measure the distinguishability of single photons from our source. But for our experimental setup, we conjecture that they should be nearly indistinguishable. Because all the vortex photon with non-Gaussian transverse mode structure is simultaneously converted into the Gaussian mode, i.e., $\{\ket{-1},\ket{0},\ket{1}\}->\{\ket{0},\ket{0},\ket{0}\}$, we can't distinguish definitely where the photon state $\ket{0}$ comes from each of OAM photon pair, which ensures the distinguishability in spatial DOF. The HOM interferences for single OAM photon pair ensure they are identical in the frequency domain. Also, the multiplexed single photon source is indistinguishability in DOFs of temporal and polarization because of the same optical path and polarization. }}
\begin{table*}[tph]
  \centering
  \begin{tabular}{|c|c|c|c|c|}
 \hline
Methods& OAM-mode& Efficiency & Advantages & Ref.\\
 \hline
Interferometer with Dove lens& Odd and even mode & 100\% in principle & High-sorting visibility & \cite{leach2002measuring,leach2004interferometric}\\
 \hline
 Forked diffraction blazing & Arbitrary OAM mode & Linear losses of 1/N & The less optical elements& \cite{gibson2004free,Yao2011a}\\
 \hline
 Log-polar coordinate transformation& Arbitrary OAM  mode & 100\% in principle &The high-sorting visibility & \cite{berkhout2010efficient, mirhosseini2013efficient,ruffato2017compact}\\
  \hline
\end{tabular}
  \caption{The main methods for OAM mode sorting. }
\end{table*}
\section{Discussion}
We experimentally investigate an HSPS by multiplexing the multi-mode of down-converted photon pairs emitted from the nonlinear crystal. The photon rate of OAM-HSPS obtained a considerable enhancement of 47\% by multiplexing three OAM modes. A small second-order autocorrelation function value $g^{(2)}(0)$ illustrates that the generated photons are with high single photon purity. The enhancement can be improved in further if we decrease the overall losses in the system. The losses of the system, mainly caused by the OAM sorting and converting method employed in our experiment, limit the overall HSPS rate. Nevertheless, the overall losses do not affect the enhancement of multiplexing OAM modes in principle.

In the future, we plan to conduct some significant explorations of the demonstrated protocol.  On the one hand, our proof-of-principle OAM-HSPS shows that we can construct a good HSPS by multiplexing more OAM modes, i.e., by multiplexing high-dimensional entangled state in OAM space \cite{Dada2011}, or high-dimensional maximally entangled state \cite{vaziri2002experimental,liu2018coherent, kovlakov2018quantum}. Recently, we realized an arbitrary superposition state of even 50 OAM modes, an analogous OAM-Schr$\rm \ddot{\textup{o}}$dinger cat state in Ref \cite{liu2018generation}, so this platform for OAM-HSPS is with excellent scalability.  On the other hand, the theoretical protocol proposed in this article can be directly realized in some new micro- materials supporting the spatial modes, i.e., the photonic chip \cite{feng2016chip,chen2018mapping}, where we can reduce the losses in mode- coupling and sorting.

We believe that our works can be considered an important first step towards reaching a high-quality and well-scalability HSPS. The unsolved challenges are to implement a lossless OAM-sorter working with as more as possible OAM modes. Because the manipulation of a lossless OAM-sorter is still a well-defined problem, it has a great potential for the development of a highly efficient OAM-based HSPS.

\section{Acknowledgments}
This work is partially supported the Anhui Initiative in Quantum Information Technologies (AHY020200);
National Key R$\&$D Program of China (2018YFA0307400); National
Natural Science Foundation of China (61435011, 61525504, 61605194, 61775025, 61405030); China Postdoctoral Science Foundation (2016M590570, 2017M622003); Fundamental Research Funds for the Central Universities.

\appendix
\section{The methods for sorting OAM modes}

The biggest advantages for OAM-HSPS is well scalable in dimension.
However, it is still a challenge to realize a perfect OAM sorter for many OAM modes. Various OAM sorters have been made to reduce the losses and mode cross talk. We listed some mainstream methods in  table 1 to construct an OAM sorter.

 The methods used in our manuscript is the 'Forked diffraction blazing ', where the advantage is of only one optical element (SLM), while the sorting-efficiency will reduce around 1/N. For two other methods in the above table, the sorted efficiency can be improved in principle, while the most significant disadvantage is that there needs many optical elements, i.e., three SLMS needs in the method of Log-Polar transformation, which increase the overall losses inevitably.
 The quality of OAM-based HSPS will be improved by reducing the losses further by using more efficient schemes in the future.
 
%\bibliography{OAM_multiplexing}

\end{document}